\documentclass{article}
\usepackage{spconf,amsmath,amssymb,graphicx,hyperref}
\usepackage{booktabs}
\usepackage{threeparttable}

\usepackage{tikz}
\usepackage{atbegshi}
\usepackage{ragged2e} 

\AtBeginShipoutFirst{%
  \begin{tikzpicture}[remember picture,overlay]
    \node[anchor=north,yshift=-10pt] at (current page.north) {%
      \fbox{%
        \begin{minipage}{0.80\paperwidth} 
          \scriptsize
          \noindent
          \justifying
          © 2026 IEEE. Personal use of this material is permitted. Permission from IEEE must be obtained for all other uses, in any current or future media, including reprinting/republishing this material for advertising or promotional purposes, creating new collective works, for resale or redistribution to servers or lists, or reuse of any copyrighted component of this work in other works.
        \end{minipage}%
      }%
    };
  \end{tikzpicture}%
}

\title{MMED: A Multimodal Micro-Expression Dataset based on Audio-Visual Fusion}
\name{Junbo Wang \qquad Yan Zhao$^{\star}$\thanks{$^{\star}$Corresponding author. This research is supported by the National Natural Science Foundation of China under Grant 62571215 and 62271226 and Jilin Provincial Science and Technology Development Plan Project under Grant 20250102208JC.} \qquad Shuo Li \qquad Shibo Wang \qquad Shigang Wang \qquad Jian Wei}
  
\address{College of Communication Engineering, Jilin University}

\begin{document}
\ninept
\maketitle
\begin{abstract}
Micro-expressions (MEs) are crucial leakages of concealed emotion, yet their study has been constrained by a reliance on silent, visual-only data. To solve this issue, we introduce two principal contributions. First, MMED, to our knowledge, is the first dataset capturing the spontaneous vocal cues that co-occur with MEs in ecologically valid, high-stakes interactions. Second, the Asymmetric Multimodal Fusion Network (AMF-Net) is a novel method that effectively fuses a global visual summary with a dynamic audio sequence via an asymmetric cross-attention framework. Rigorous Leave-One-Subject-Out Cross-Validation (LOSO-CV) experiments validate our approach, providing conclusive evidence that audio offers critical, disambiguating information for ME analysis. Collectively, the MMED dataset and our AMF-Net method provide valuable resources and a validated analytical approach for micro-expression recognition.
\end{abstract}
\section{Introduction}
\label{sec:intro}

In the procedure of human communication, micro-expressions (MEs) serve as fleeting yet potent signals of a person's true emotional state, often revealing suppressed feelings in scenarios ranging from courtrooms to high-stakes games of social deception\cite{10.1093/acprof:oso/9780195327939.003.0008}. Despite their importance in fields like national security and clinical psychology\cite{9915437}, the automatic recognition of MEs remains a formidable challenge, deeply intertwined with the limitations of available data.

To this end, the research community has made considerable strides in advancing MER, primarily by refining visual analysis through sophisticated feature learning\cite{fang2024spcl, wang2024micro}, tackling data scarcity with strategies spanning from knowledge transfer\cite{wang2024progressively} to advanced augmentation\cite{gu2024facial, chen2025improving}, and broadening the scope of the task itself\cite{zou2025synergistic}. The prevailing focus of these advancements has been on leveraging unimodal visual data, with many state-of-the-art approaches dedicated to fusing different facets of visual information, such as RGB frames with optical flow\cite{ma2025multi, xie2025micro}. In parallel, the question of how genuinely distinct sensory channels, such as the auditory modality, might contribute to this task remains a less explored but equally compelling area of inquiry.

Early datasets often featured posed expressions, with pioneering work by Polikovsky et al.\cite{5522282} serving as a key example, while subsequent work progressed towards eliciting spontaneous reactions in controlled laboratory settings. Further advancing this trajectory, benchmarks like MEVIEW\cite{MEVIEW} introduced "in-the-wild" data, and CAS(ME)³\cite{9774929} pioneered the use of a simulated crime scenario to increase ecological validity. In parallel with these efforts to improve realism, the potential of multimodal data is also explored. Early explorations, such as SMIC (RGB+NIR)\cite{6126401} and 4DME (4D data)\cite{9796028}, primarily focused on fusing different facets within the broader visual domain. The CAS(ME)³\cite{9774929} dataset notably advanced the exploration of multimodal data, primarily by demonstrating the value of depth information for enhancing MER performance. In parallel with this finding, the dataset also incorporated an audio modality. However, as the authors themselves noted, their initial exploration yielded unsatisfactory results, which they attributed to challenges in signal processing and feature representation. They positioned their work as a foundational data platform, calling for further research to optimize the processing of speech signals and fully explore the modality's potential impact on ME analysis\cite{9774929}.

Psychological studies have shown that emotion perception is an inherently multi-modal process, where visual cues are naturally integrated with auditory information—such as shifts in vocal tone, breathing patterns, or subtle non-verbal sounds—to form a holistic understanding\cite{seeing}. Thus, a resource that can enable a deep exploration of audio-visual synergy in micro-expressions is required. Based on this, we generate a publicly available audio-visual micro-expression dataset captured in a realistic setting: MMED. Our main contributions are summarized as follows:

1) A Novel Audio-Visual Micro-Expression Dataset. We present MMED, to our knowledge, the first dataset capturing synchronized audio-visual recordings of micro-expressions elicited from real-world interactions. This dataset serves as a practical tool to enable micro-expression recognition by leveraging both auditory information and traditional visual cues.

2) A Strong Multi-Modal Fusion Baseline. To validate the effectiveness of our dataset, we propose a novel Asymmetric Multimodal Fusion Network (AMF-Net), which introduces an asymmetric cross-attention mechanism to effectively fuse global visual summaries with dynamic temporal audio cues, establishing a strong performance benchmark and offering a viable path for future multi-modal ME analysis.

3) Empirical Validation of the Audio Modality. Through comprehensive experiments, we validate the effectiveness of our proposed dataset and fusion network. Our findings show that multi-modal fusion significantly outperforms vision-only approaches, validating the importance of the audio channel.

\section{MMED}
\label{sec:pagestyle}

The highly transient and low-amplitude nature of micro-expressions presents a fundamental challenge: visual signals alone can be ambiguous. The auditory channel, however, offers a rich source of disambiguating information, where cues such as vocal prosody and pitch can provide the crucial emotional context needed to interpret subtle facial movements\cite{PAN20211}. To enable the systematic exploration of this audio-visual synergy, we constructed the Multimodal Audio-Visual Micro-Expression Dataset (MMED). The design and construction of which are detailed in the following sections.

\subsection{Rationale and Data Sourcing}
\label{ssec:subhead}

A primary goal in generating MMED is to address the ecological validity gap that exists in many existing datasets\cite{s22041524, oh2018survey}. To achieve this, we moved beyond laboratory settings and sourced our data from a high-stakes, socially interactive environment: online "Werewolf" game competitions. In this scenario, players possess a strong desire to win and must manage their social presentation under pressure, making their emotional experiences and suppression attempts more authentic. Unlike paradigms that rely on passive responses to stimuli, our approach captures micro-expressions as they are naturally triggered during genuine verbal communication and social deduction, providing a dataset with high ecological validity. Furthermore, the game requires each participant to deliver speeches in a high-pressure context, which ensures a rich source of corresponding audio signals.

\subsection{Annotation Protocol}
\label{ssec:subhead}

As noted by Li et al.\cite{9774929}, annotating MEs is an exceptionally laborious and time-consuming endeavor. The transient and subtle nature of these expressions makes their detection significantly more challenging than that of macro-expressions. To ensure the highest quality of annotation, we implemented a multi-stage protocol:

1) Initial Screening: All video footage is first reviewed to identify and isolate segments containing potential micro-expressions. This crucial first pass serves to filter out the majority of emotionally irrelevant facial movements.

2) Expert Verification: The shortlisted segments are then subjected to a rigorous review by FACS-certified expert. The expert identifies true micro-expression events and meticulously marks the onset, apex, and offset frames. Action Units (AUs) are coded, and an emotion category is assigned to each validated ME. Based on the existing criteria\cite{time}, the micro-expression we defined is within the total duration not exceeding 500ms.

3) Inter-Annotator Agreement (IAA): To ensure reliability, the entire dataset is independently annotated a second time by a different annotator. The inter-annotator agreement score is then calculated using:

\begin{equation}
\begin{split}
r=\frac{2×\left | A1\bigcap A2 \right | }{\left | A1 \right |+\left | A2 \right | }
\label{eq:r}
\end{split}
\end{equation}
where $\left | A1\bigcap A2 \right |$ is the number of AUs on which both annotators agreed, and $\left | A1 \right | +\left | A2 \right |$ is the total number of AUs coded by each. The resulting agreement score of 0.89 indicates a high degree of reliability in our annotations. Any remaining discrepancies are resolved through discussion to form the final, consensus-based annotation file.

For emotion categorization, our protocol adapts the 4DME mapping\cite{9796028}, but with a fitment modification: the exclusion of the 'repression' category, which stems from the characteristic of our dataset—the near-universal presence of AU50 (mouth opening) in our audio-visual clips. The prevalence of this AU is particularly consequential for the 'repression' category, as scoring an associated AU in a conversational context requires it to persist for at least one syllable or have its apex coincide with a speech pause. We empirically found that the vast majority of AU50 instances in our dataset did not satisfy this temporal criterion. To maintain annotation rigor and avoid mislabeling, we therefore omitted the 'repression' category. Our final annotation scheme is thus comprised of four categories: Positive, Negative, Surprise, and Others, with the 'Others' category for emotionally ambiguous signals.

\subsection{Dataset Statisticsl}
\label{ssec:subhead}

As illustrated in Figure 1, the final annotated dataset exhibits a class imbalance, with a higher percentage of "Surprise" and "Positive" samples compared to the "Negative" category. Far from being a limitation, this distribution is a common characteristic of spontaneous emotion datasets, reflecting the natural frequency of emotional occurrences and the inherent challenges in eliciting them\cite{oh2018survey}. We therefore consider it a realistic feature of the dataset that provides a valuable challenge for developing robust MER algorithms.

\begin{figure}[htb]
  \centering
  \begin{minipage}[b]{\linewidth}
    \centering
    \includegraphics[width=0.95\linewidth]{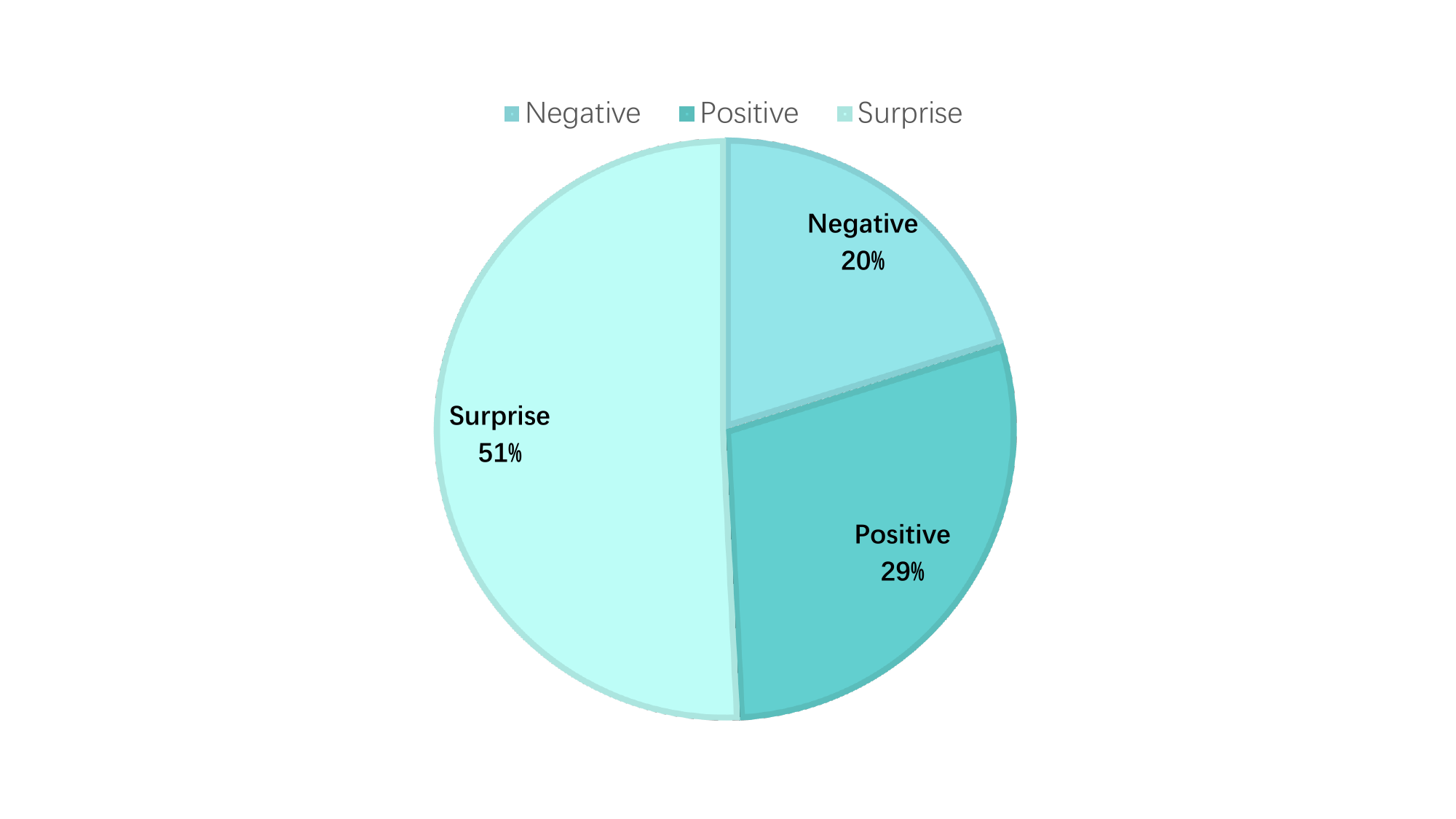}
    \centerline{(a) Three-class ME}\medskip
  \end{minipage}
  \begin{minipage}[b]{\linewidth}
    \centering
    \includegraphics[width=0.95\linewidth]{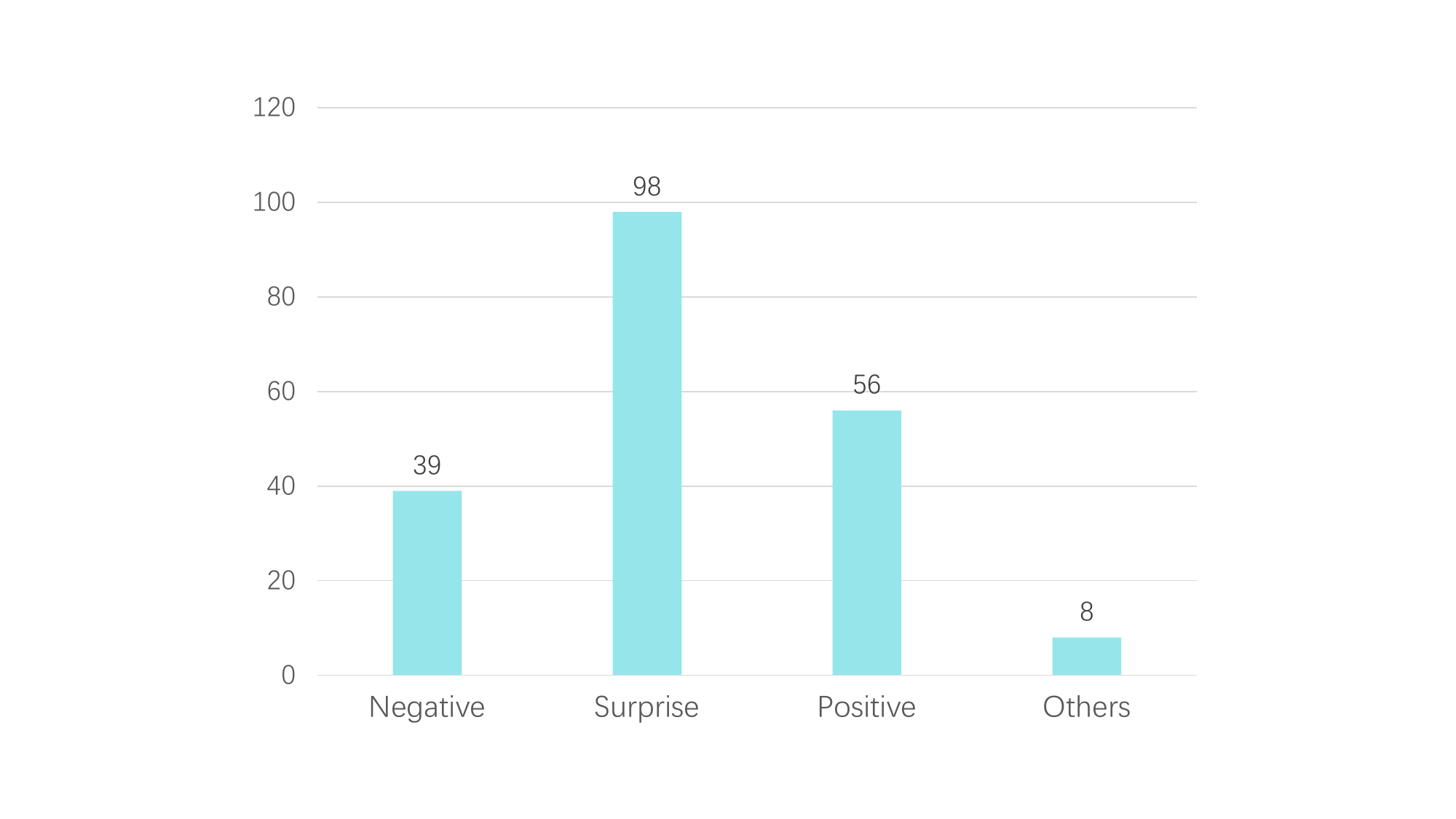}
    \centerline{(b) Overview of the ME Dataset}\medskip
  \end{minipage}
  
  \caption{Statistical analysis of annotated samples in the MMED.}
  \label{fig:vertical_figures}
\end{figure}

\section{Asymmetric Multimodal Fusion Network}
\label{sec:typestyle}

\begin{figure*}[htb]
  \centering
  \includegraphics[width=\linewidth]{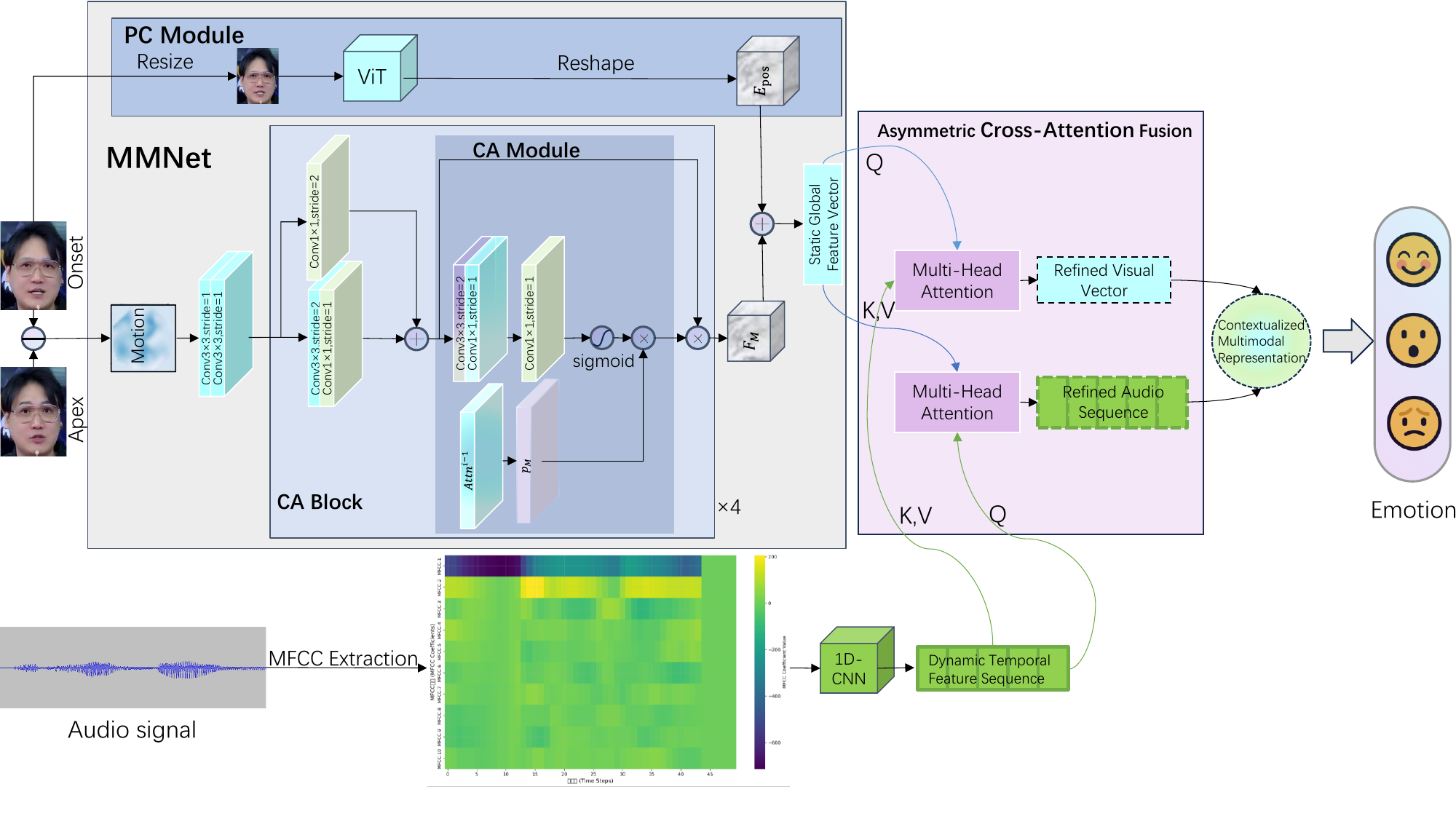}
  \caption{The Asymmetric Multimodal Fusion Network (AMF-Net).}
  \label{fig:AMF-Net}
\end{figure*}

To leverage the complementary nature of audio-visual signals in MER, we propose the Asymmetric Multimodal Fusion Network (AMF-Net). Illustrated in Figure 2, AMF-Net is engineered around a core principle: fusing features that are true to the intrinsic properties of each modality. It therefore employs an asymmetric two-branch architecture. The visual branch uses an MMNet\cite{ijcai2022p150} backbone to distill a micro-expression video into a global feature vector, encapsulating the holistic facial movement. The audio branch utilizes a 1D-CNN to preserve the temporal evolution of acoustic cues, outputting a dynamic feature sequence.

The key to AMF-Net is its fusion module, designed to resolve the fundamental challenge of integrating these heterogeneous representations. Without using traditional concatenation, we employ an asymmetric cross-attention mechanism to enable a deep, bi-directional contextualization between the modalities. The process is as follows:

1)Feature Projection: Initially, to ensure dimensional compatibility for the attention mechanism, the visual vector and the audio sequence are projected into a common latent space of dimension $D_{c}$ using separate linear layers.

2)Cross-Attention Mechanism: Two multi-head cross-attention modules are arranged in parallel, following the standard Scaled Dot-Product Attention paradigm. For a single attention head, the operation is defined as:

\begin{equation}
\begin{split}
\mathrm{head}_{i}
&= \operatorname{Attention}\bigl(QW_{i}^{Q},KW_{i}^{K},VW_{i}^{V}\bigr)\\
&= \operatorname{softmax}\!\left(\frac{(QW_{i}^{Q})(KW_{i}^{K})^{T}}{\sqrt{d_{k}}}\right)(VW_{i}^{V}).
\label{eq:head}
\end{split}
\end{equation}
where ($W_{i}^{Q}$,$W_{i}^{K}$,$W_{i}^{V}$) is the learnable projection matrix at the $i$th head, and $d_{k}$ is the dimension of the key vector. The final output of the Multi-Head Attention layer is obtained by concatenating the outputs of all heads, which is then passed through a final linear projection. This is formally expressed as:

\begin{equation}
\begin{split}
\operatorname{MultiHead}(Q,K,V)
&= \operatorname{Concat}\bigl(\mathrm{head}_{1}, \ldots, \mathrm{head}_{h}\bigr) W^{O}
\label{eq:multihead}
\end{split}
\end{equation}
where $h$ is the number of parallel attention heads. The Concat operation joins the output vectors from each head. The matrix $W^{O}$ is the output projection matrix, a learnable parameter that transforms the concatenated feature vector back into the model's expected dimension, effectively mixing and synthesizing the information learned from all subspaces.

This mechanism is deployed in two parallel, complementary streams: 1) Audio Refines the Visual Summary (VA-Attention). In this stream, the visual vector acts as the query, effectively "interrogating" the entire audio sequence to find the most relevant acoustic cues that can enrich its own representation. This allows the model to answer the question: "Given this overall facial expression, which parts of the accompanying sound are the best match?"

\begin{equation}
\begin{split}
\tilde{\mathbf{v}}_{\text{seq}}
= \operatorname{MultiHead}\!\left(\mathbf{v'}_{\text{seq}}, \mathbf{A'}, \mathbf{A'}\right)
\in \mathbb{R}^{1 \times D_c}
\label{eq:va_attention}
\end{split}
\end{equation}
Here, the Query is the projected visual vector (Q=${V}'_{seq}$), while the Keys and Values are derived from the projected audio sequence (K=A=${A}'$). The output,$\tilde{V_{seq}}$, is a contextually enriched visual vector, now informed by the temporal dynamics of the audio.

2)Visual Grounds the Audio Sequence (AV-Attention). Conversely, each temporal step in the audio sequence acts as a query to "consult" the visual vector. This allows the model to answer the question: "How should each moment of this audio signal be interpreted, given the stable context of the overarching facial expression?"

\begin{equation}
\tilde{\mathbf{A}}
= \operatorname{MultiHead}\!\left(\mathbf{A'}, \mathbf{v'}_{\text{seq}}, \mathbf{v'}_{\text{seq}}\right)
\in \mathbb{R}^{T_a \times D_c}
\label{eq:av_attention}
\end{equation}
Here, the Query is the projected audio sequence (Q=${A}'$), while the Keys and Values are derived from the visual vector (K=V=${V}'_{seq}$). The result, $\tilde{A}$, is a new audio sequence where each element has been contextualized and modulated by the global visual information.

Classification Head: The resulting fused representation, which now encodes contextually enriched audio-visual information, is passed through a pooling layer and a final classifier to predict the micro-expression category.

\section{Experiments}
\label{sec:majhead}

To ensure a rigorous, person-independent evaluation, all experiments follow the LOSO-CV protocol. For the metrics, we provide both Accuracy (Acc) and the Unweighted F1-score (UF1), the latter of which is crucial for assessing performance on imbalanced data. For fair comparison, all baseline methods follow the hyperparameter settings described in their original publications.

\subsection{MER based on Visual Information}
\label{ssec:subhead}

To select a strong visual backbone and concurrently validate the effectiveness of the visual information within MMED, we began by benchmarking several state-of-the-art MER methods in a visual-only setting. For this unimodal experiment, the audio channel of the dataset is intentionally excluded. We implemented a range of representative models, including KFC-MER\cite{9428407}, MMNet\cite{ijcai2022p150}, and HTNet\cite{WANG2024128196}, to identify the most effective architecture for subsequent multimodal integration.

The results, summarized in Table 1, affirm the validity of MMED as a dataset for visual analysis. The strong performance of leading methods like HTNet\cite{WANG2024128196} and MMNet\cite{ijcai2022p150}, which achieved accuracies approaching 80\%, indicates that MMED contains rich, discriminative, and learnable visual features. Furthermore, the relative performance ranking of these models on MMED is highly consistent with their rankings on established benchmarks such as SMIC\cite{6126401}, CASME II\cite{Yan2014CASMEII}, and SAMM\cite{7492264}, confirming the reliability of our dataset for comparative evaluation. This consistency is particularly crucial, as we did observe that most methods exhibited slightly lower absolute scores on MMED compared to other benchmarks (see Table 2). We attribute this not to a lack of quality, but to the unique challenges our dataset introduces: the co-occurrence of MEs with speech can lead to associated mouth movements that mask or interfere with subtle muscle activations. This presents a more difficult and realistic recognition scenario. Therefore, the consistent performance hierarchy allowed us to confidently select MMNet, the top-performing model under these challenging conditions, as a robust visual backbone for our subsequent multi-modal experiments.

\begin{table}[!ht]
    \centering
    \caption{Three-class classification performance comparison of baseline MER methods on MMED dataset without using audio sequence}
    \label{tab:performance_comparison}
    \vspace{1pt} 
    \begin{tabular}{lll}
    \hline
        Method & Acc (\%) & UF1 \\ \hline
        KFC-MER\cite{9428407} & 68.06 & 0.6335 \\ 
        MMNet\cite{ijcai2022p150} & \textbf{78.54} & 0.7057 \\ 
        HTNet\cite{WANG2024128196} & 78.01 & \textbf{0.7494} \\ \hline
    \end{tabular}
\end{table}

\begin{table}[t]
\centering
\caption{Three-class classification performance comparison of baseline MER methods on datasets SMIC, CASME II and SAMM}
\vspace{1pt}
\resizebox{\linewidth}{!}{
\label{tab:mer_results}
\begin{threeparttable}
\begin{tabular}{lcccccccc}
\toprule
\multicolumn{1}{c}{Method} & \multicolumn{2}{c}{SMIC} & \multicolumn{2}{c}{CASME II} & \multicolumn{2}{c}{SAMM} \\
\cmidrule(lr){2-3}\cmidrule(lr){4-5}\cmidrule(lr){6-7}\cmidrule(lr){8-9}
 & Acc (\%) & UF1 & Acc (\%) & UF1 & Acc (\%) & UF1 \\
\midrule
KFC-MER\cite{9428407} & 65.85 & 0.6638 & - & - & - & - \\
MMNet\cite{ijcai2022p150} & - & - & 95.51 & 0.9494 & 90.22 & 0.8391 \\
HTNet\cite{WANG2024128196} & - & 0.8049 & - & 0.9532 & - & 0.8131 \\
\bottomrule
\end{tabular}
\begin{tablenotes}[flushleft]
\footnotesize
\item The ’-’ in the table indicates that the information is not provided in the original paper.
\end{tablenotes}
\end{threeparttable}
}
\end{table}

\begin{table}[!ht]
    \centering
    \caption{Performance metrics of different modalities on MMED dataset}
    \label{tab:performance_comparison}
    \vspace{1pt} 
    \begin{tabular}{lll}
    \hline
        Modality Type & Acc (\%) & UF1 \\ \hline
        Visual & 78.54 & 0.7057 \\ 
        Audio & 75.16 & 0.6914 \\ 
        Visual+ Audio & \textbf{81.90} & \textbf{0.7060} \\ \hline
    \end{tabular}
\end{table}

\subsection{MER based on Multimodal Fusion}
\label{ssec:subhead}

Table 3 quantifies the empirical value of the audio modality in MER by directly comparing the performance of three settings under a rigorous LOSO-CV protocol: Audio-Only, Visual-Only, and Visual + Audio (fused by AMF-Net). The results show several key insights:

1)Audio modality contains discriminative emotional cues. The audio-only model achieves a notable 75.16\% accuracy and 0.6914 UF1 score on its own. This finding is significant, as it provides the first quantitative evidence that the non-verbal acoustic cues co-occurring with MEs are more than just background noise, they carry discriminative, emotion-related information that a model can successfully learn. 2) Multi-modal fusion yields substantial performance gains. The fused AMF-Net (Visual + Audio) model achieves a significant improvement in performance. It obtains an absolute gain of 3.36\% in accuracy over the strong visual-only baseline and an even more substantial 6.74\% gain over the audio-only baseline.3) Visual and audio modalities are complementary. The synergistic result strongly suggests that the visual and audio channels offer complementary, rather than redundant, information. The fusion model can leverage cues from one modality to resolve ambiguities present in the other, leading to a more robust and accurate classification. This demonstrates the clear advantage of a multi-modal approach for the MER task.

\section{Conclusions}
\label{sec:print}

In this paper, we first introduce MMED, to our knowledge, the first publicly available audio-visual ME dataset captured in an ecologically valid setting. Second, to establish a strong baseline on this new resource, we propose the AMF-Net, an effective approach for integrating global visual features with dynamic temporal audio cues. Our experiments confirm the presence of discriminative acoustic cues in MEs (75.16\% audio-only accuracy) and demonstrate that their fusion with visual data via AMF-Net significantly boosts performance to 81.90\%. A critical analysis of the Unweighted F1-score, however, reveals that this gain primarily benefits the majority classes. This finding indicates that the fundamental challenge of class imbalance remains largely unaddressed by the fusion process itself. These findings direct future work towards addressing this imbalance, potentially through few-shot learning or advanced data augmentation strategies. We also plan to explore the integration of additional modalities, such as transcribed linguistic content, to develop a more comprehensive understanding of MEs.



\vfill\pagebreak


\bibliographystyle{IEEEbib}
\bibliography{strings,refs}

\end{document}